\documentclass[twocolumn,showpacs,eqsecnum,aps,prd,amsfonts,amssymb]{revtex4}

\usepackage{graphicx}% Include figure files
\usepackage{bm}% bold math
\usepackage{amsmath}

\DeclareMathOperator{\sech}{sech}
\def\bra#1{\langle #1 |}
\def\ket#1{| #1 \rangle}

\begin{document}

\title{Detection of  negative energy:  4-dimensional examples}

\author{P.C.W. Davies}
\affiliation{Department of Physics and Mathematical Physics,
    University of Adelaide, Box 498 G.P.O.,  Adelaide, S.A. 5001, Australia} 

\author{Adrian C. Ottewill} 
\affiliation{Department of Mathematical Physics, University College Dublin,
Belfield, Dublin 4, Ireland}
\email{ottewill@relativity.ucd.ie}

%\draft
\begin{abstract}
We study the response of switched particle detectors to static
negative energy densities and negative energy fluxes.  
It is demonstrated how the switching leads to excitation
even in the vacuum and how negative energy can lead to a suppression
of this excitation.  
We obtain quantum inequalities on the detection similar to
those obtained for the energy density by Ford and co-workers and in an
`operational' context by Helfer. 
We revisit the question `Is there a quantum equivalence principle?'
in terms of our model.  Finally, we briefly address the issue
of negative energy and the second law of thermodynamics.
\end{abstract}

\pacs{04.62.+v, 03.70.+k}

\maketitle

\section{Introduction}
\label{sec:intro}

While in classical physics the energy density of a field is strictly
positive, quantum field theory allows states containing regions of
negative energy density or negative energy fluxes~\cite{Epstein:1965}.  The
Casimir vacuum between two conducting plates and squeezed states
provide two familiar examples of such states, both of which have been
studied experimentally.  In these regions of
negative energy density the standard `local energy conditions' assumed
in classical general relativity no longer hold. This gives rise to the
possibility of avoiding the theorems of classical general relativity
such as the singularity theorems and the Hawking black hole area
increase theorem thus allowing for black hole evaporation to occur.
Recent interest in these states has surrounded the apparent violation
of cherished beliefs that such states might entail, by appearing to allow
the existence of traversable wormholes and `time 
machines'~\cite{Morris:1988a,Morris:1988b} and
violations of cosmic censorship~\cite{FordRoman:1990,FordRoman:1992} 
and the second law of thermodynamics~\cite{Ford:1978,Davies:1982}. 

One powerful approach that  has evolved to prove that any such violations 
are limited to microscopic fluctuations and cannot produce any macroscopic
effect is that of quantum inequalities constraining the magnitude and
duration of negative energy
regions~\cite{Ford:1978,Ford:1991,FordRoman:1995,FordRoman:1997,Ford:1998,Ford:1998a,
FewsterEveson:1998}.
 An example of such a result in four-dimensional
space-time is due to Ford and Roman~\cite{FordRoman:1995} who showed
that  for a free massless scalar field in Minkowski space-time
\begin{equation}
\label{FordRomanInequality}
 \int_{-\infty}^{\infty}  \lambda^2(t) \langle \hat \rho(t) \rangle 
\textrm{d}t \geq - \frac{3}{32 \pi^2 T{}^4}
\end{equation}
where $\langle \hat \rho(t) \rangle$ is the expectation value of 
the energy density (in the frame of an arbitrary inertial observer
whose time coordinate is $t$) in an arbitrary quantum state, and 
\[
  \lambda^2(t) = \frac{T}{\pi} \frac{1}{t^2 + T^2} 
\]
is a `sampling function' with characteristic width $T$. 
For general $\lambda(t)$, Fewster and Eveson~\cite{FewsterEveson:1998}
proved the stronger and more general bound
\begin{equation}
\label{FewsterEvesonInequality}
 \int_{-\infty}^{\infty}  \lambda^2(t) \langle \hat \rho(t) \rangle 
\textrm{d}t \geq - \frac{1}{16 \pi^2}  \int_{-\infty}^{\infty}
\bigl(\lambda''(t)\bigr)^2 \textrm{d}t , 
\end{equation}
which was extended to static space-times by Fewster and
Teo~\cite{FewsterTeo:1999}.
(Stronger, indeed optimal, results have been proved in unbounded 
two-dimensional space-time~\cite{Flanagan:1997} but 
the physics of particle detection is very different in this case and
we choose to discuss it elsewhere.)

Central to the issue of negative energy and the second law of
thermodynamics is the question of how atoms respond to negative energy
fluxes.  The issue is particularly subtle since the quantum
inequalities indicate some restriction on the length of time for which
a significant negative energy flux can be sustained.  As a result one
finds that in discussions of the detection of negative energy fluxes
one comes face to face with the infamous $\Delta E \Delta t$
uncertainty principle: If one has an unswitched detector then the
effects of the negative energy flux are swamped by the positive energy
which must surround it.  If one tries to measure whether the detector
is excited or not while the flux is passing through then that
measurement must be made so fast that the switching itself necessarily
excites the detector.  These issues were bravely tackled by
Grove~\cite{Grove:1988}, however, his results are clouded by the
complications of his analysis, and by the non-standard coupling that
he chooses.  The issues were further addressed  by Ford et
al~\cite{Ford:1992} who studied the response of an array of
quantum-mechanical spin-$\frac12$ particles to negative energy fluxes.  
The role of the $\Delta E \Delta t$
uncertainty principle was again emphasied by
Helfer~\cite{Helfer:1998} who formulated an `operational' energy
condition on the basis of it: ``the energy of an isolated device
constructed to measure or trap the energy in a region, plus the energy
it measures or traps, cannot be negative.''

In this paper we shall examine the response of `particle detectors' to
negative energy fluxes.  To be able to concentrate our measurements on
periods of negative energy flux we explicitly switch our detector on
and off.  This introduces excitations even in the vacuum which we
discuss in some detail. To isolate the effects of the negative energy
we then compare the response of a detector switched on and off during
a period of negative energy density (or negative energy flux) and that
switched on and off in the vacuum.  We show that, in line with Grove's
two-dimensional results, the negative energy can lead to a suppression
of excitations that would have occurred for the detector in the
vacuum. However, we additionally show that there exists a quantum
inequality limiting the size of this effect.

Our analysis also enables us to revisit the question of the response
of an inertial detector moving through the Rindler vacuum.  This
situation was originally studied by Candelas and
Sciama~\cite{Candelas:1984}.  These authors considered a particular
limit where the observation time went to infinity while the final
acceleration remained fixed and found that, in this limit, the
detector did not respond to the negative energy density of the Rindler
vacuum but instead responded just as if it were in the Minkowski vacuum.
While our analysis confirms this result it also shows that there are
interesting effects of the Rindler negative energy which are simply
lost in this limit.

 In this paper we have concentrated purely on four-dimensional
examples, leaving the many interesting two-dimensional examples to a
separate publication.  The principal reason for this is that in two
dimensions there are mathematical and physical reasons for preferring a
coupling to $\dot \varphi$ rather than $\varphi$ (arising from the
poor infrared behavior of the massless theory); as a
coupling to $\varphi$ is more conventional in the four-dimensional literature we
prefer to use it here.  In addition, as mentioned above, the quantum
inequalities which have been proved vary between two and four
dimensions with much tighter results available in two-dimensional
space-time~\cite{Flanagan:1997}.

We set $\hbar=c=1$ and use the space-time conventions of
\cite{MTW:1973}.

\section{The model}
\label{sec:model}

We shall deal exclusively with a real scalar field, $\varphi$, and
since the effects of negative energy are most pronounced for massless
fields we shall restrict ourselves to that case.  Our model is a
simple generalization of the standard monopole detector in which we
include an explicit switching factor.  Thus we shall write our
interaction Lagrangian as
\begin{equation}
 \int {\rm d}\tau \> \lambda(\tau) m(\tau) \varphi (x(\tau))
 \label{eq:2.1}
\end{equation}
where $\tau$ denotes proper time along the world-line of the detector,
$m(\tau)$ denotes the monopole moment of the detector and
$\lambda(\tau)$ is an real switching factor which we have introduced
so that we can make measurements over restricted time intervals.  We
assume that the evolution of the monopole is determined by a
time-independent Hamiltonian, $\hat H_D$, and that the monopole has
corresponding energy eigenstates, which we may denote by $\ket{E_1}$
and $\ket{E_2}$.  Working in the interaction picture the monopole
moment then evolves in the standard fashion
\[
   \hat m(\tau)= {\rm e}^{i\hat H_D \tau} \hat m (0) {\rm e}^{-i \hat
                 H_D \tau} .
\]

If the field is initially in the state $\ket{A}$ then by standard
first-order perturbation theory we obtain the probability for a transition
between the two states of the detector as
\begin{equation}
 P_A(E_1 \to E_2) = \bigl|\bra{E_1} \hat m(0) \ket{E_2} \bigr|^2 \,
       \Pi_A(E_2 - E_1)
\label{eq:2.3}
\end{equation}
where
\begin{eqnarray}
    \Pi_A(E) \equiv \int\limits_{-\infty}^{\infty} {\rm d}\tau\>
       \int\limits_{-\infty}^{\infty} {\rm d}\tau'\> {\rm
       e}^{-iE(\tau-\tau')} \lambda(\tau) \lambda(\tau') \nonumber \\
        \times \bra{A}\hat\varphi\bigl(x (\tau)\bigr)
       \hat\varphi\bigl(x(\tau')\bigr)\ket{A} .
\label{eq:2.4}
\end{eqnarray}
The prefactor in Eq.~(\ref{eq:2.3}) merely contains information about
the details of the detector, the real interest lies in the function,
$\Pi_A(E)$, defined in Eq.~(\ref{eq:2.4}).  We shall refer to $\Pi_A(E)$
as the response function.  We shall be interested in both excitations
$E > 0$ and de-excitations $E < 0$.

It is convenient to introduce the Fourier transform of the switching
function, $\tilde \lambda(\omega)$, with conventions defined by the
equation
\begin{equation}
  \tilde \lambda(\omega) = \int {\rm d}\tau\> {\rm e}^{-i\omega \tau}
  \lambda(\tau) .
\label{eq:2.5}
\end{equation}
From the reality of $\lambda(\tau)$ it follows that ${\tilde
\lambda}^*(\omega) = \tilde \lambda(- \omega)$, an equality that we
shall use freely in the following.  It is possible to isolate the
dependence on the switching from that on the state by writing
Eq.~(\ref{eq:2.4}) in the form
\begin{equation}
\Pi_A(E) = \frac{1}{4\pi^2} \int\limits_{-\infty}^{\infty} {\rm
        d}\omega\> {\tilde \lambda}(\omega)
        \int\limits_{-\infty}^{\infty} {\rm d}\omega'\> {\tilde
        \lambda}^*(\omega ') \> \pi_A(E;\omega,\omega ')
\label{eq:2.6}
\end{equation}
where
\begin{eqnarray}
  \pi(E;\omega,\omega ') \equiv \int\limits_{-\infty}^{\infty} {\rm
d}\tau\> \int\limits_{-\infty}^{\infty} {\rm d}\tau'\> {\rm
e}^{-i(E-\omega)\tau+i(E-\omega ')\tau'} \nonumber \\
\times  \bra{A}\hat\varphi\bigl(x
(\tau)\bigr) \hat\varphi\bigl(x(\tau')\bigr)\ket{A}
\label{eq:2.7}
\end{eqnarray}
is independent of the switching function, $\lambda(\tau)$.

We shall study the response under a range of switchings but we choose
them all to be functions of a single dimensionless variable
$(\tau -\tau_0)/T$ with the two parameters
$\tau_0$ and $T$ determining the time of the peak and a measure of the
duration of the switching, respectively. We shall write
\begin{equation}
    \lambda(\tau;\tau_0,T) = \Lambda \left( \frac{\tau -\tau_0}{T}\right) .
\end{equation} 
As a consequence we can write
\begin{equation}
     \tilde \lambda(\omega;\tau_0,T) = T {\rm e}^{-i\omega \tau_0}
       \tilde \Lambda(\omega T)
\label{eq:2.8}   
\end{equation}
For convenience we will also normalise our switching functions so that
their value at $\tau_0$ is $1$, that is, $\lambda(\tau_0) = \Lambda(0)
=1$.  It follows that as we let $T \to \infty$ we recover the standard
unswitched detector results, with $\lambda(\tau)=1$, $\forall \tau$
and $\tilde \lambda (\omega) = 2 \pi \delta(\omega)$.

 The simplest choice of switching is a sudden switch on and off:
\begin{subequations}
\begin{equation}
    \Lambda_S(s) = 
\begin{cases}
1 & |s| < 1  \\
0&\text{otherwise}\\
\end{cases}
\label{eq:2.9a}
\end{equation}
giving
\begin{equation}
   \tilde \Lambda_S(\omega) = 2\frac{\sin \omega}{\omega}.
                \label{eq:2.9b}
\end{equation}
\end{subequations}
As we shall see, the suddenness of this switching leads to additional
infinities, so it is also worth considering two smoother functions
\begin{subequations}
\begin{equation}
   \Lambda_W(s) = 
\begin{cases}
1 - s^2 &|s| < 1  \\
0&\text{otherwise}\\
\end{cases} 
\label{eq:2.10a}
\end{equation}
giving
\begin{equation}
   \tilde \Lambda_W(\omega) = 4 \frac{(\sin \omega - \omega \cos \omega)}{\omega^3} ,
                                 \label{eq:2.10b}
\end{equation}
\end{subequations}
and
\begin{subequations}
\begin{equation}
   \Lambda_H (s)= 
\begin{cases}
\cos^2 \left( {\pi \over 2 }s \right) & |s|  < 1  \\
0&\text{otherwise}\\
\end{cases} 
\label{eq:2.11a}
\end{equation}
giving
\begin{equation}
   \tilde \Lambda_H(\omega) = {\pi^2 \sin \omega \over \omega(\pi^2 -
         \omega^2) } .
                   \label{eq:2.11b}
\end{equation}
\end{subequations}
These choices are inspired by the theory of data windowing and are
based on the Welch and Hanning windows respectively~\cite{Hannon:1970}.

We shall consider two further choices of switching which are not of
finite duration but which still allow us to concentrate our
measurement about one instant of time.  The first is Gaussian
switching with
\begin{subequations}
\begin{equation}
\textstyle \Lambda_G(s) = \exp\left( - {1 \over 2} s^2 \right) ,
         \label{eq:2.12a}
\end{equation}
giving
\begin{equation}
\textstyle \tilde \Lambda_G(\omega) = \sqrt{2\pi} \exp\left( - {1
   \over 2} \omega^2 \right) .  \label{eq:2.12b}
\end{equation}
\end{subequations}
The second is Cauchy (or Lorentzian) switching, corresponding to the sampling
considered by Ford~\cite{Ford:1991},
\begin{subequations}
\begin{equation}
   \Lambda_C(s) ={1 \over 1 + s^2} ,
                 \label{eq:2.13a}  
\end{equation} 
with
\begin{equation}
    \tilde \Lambda_C(\omega) = \pi {\rm e}^{-|\omega|} .
    \label{eq:2.13b}
\end{equation}
\end{subequations}

We conclude this section by observing that one may attempt to
generalize the analysis of Davies, Liu and Ottewill
~\cite{Davies:1989}, to relate the response of a switched particle
detector to the energy density it moves through.  As in
Ref.~\cite{Davies:1989} we consider the difference in response between
two different states, $\ket{A}$ and $\ket{B}$, on the space-time to
avoid problems of renormalization.  For a general motion it is
immediate from Eq.~(\ref{eq:2.4}) that
\begin{align}
\int\limits_{-\infty}^\infty {\rm d}\tau \> \lambda^2(\tau)
&\!\!\bigl\{ \bra{B} \hat\varphi^2 \left(x(\tau)\right) \ket{B} -
 \bra{A}\hat \varphi^2 \left(x(\tau)\right)\ket{A} \bigr\} =\nonumber
 \\ 
& {1 \over 2\pi} \int\limits_{-\infty}^{\infty} {\rm d}E\> \bigl\{
 \Pi_B(E) - \Pi_A(E) \bigr\} . \label{eq:2.14}
\end{align}
Eq.~(\ref{eq:2.14}) shows the close relationship between the detector response
and the average value of $\langle \hat\varphi^2 \rangle$.  In
particular, as the left hand side can be negative even when $\ket{A}$
is the vacuum state, so the right hand side must be.  In other words,
if $:$ denotes normal ordering with respect to the vacuum, then on
average the response of the detector in regions where $\bra{B} \!:
\hat \varphi^2 \! : \! \ket{B}$ is negative will be less that it would
be in the vacuum.  This is a clear four-dimensional analogue of
Grove's conclusion for two-dimensional motion, appropriate for our
more conventional choice of coupling.

The relation to the energy density is rather more tenuous.  If we
restrict ourselves to an inertial detector in Minkowski space-time
then following the methods of Ref.~\cite{Davies:1989} one can show that
\begin{align}
 &\int\limits_{-\infty}^\infty {\rm d}t \> \lambda^2(t) \bigl\{
  \bra{B}\hat \rho_\xi (t;\vec x\,)\ket{B} - \bra{A}\hat \rho_\xi
  (t;\vec x\,)\ket{A} \bigr\} = \nonumber \\ 
&{1 \over 2\pi}
  \int\limits_{-\infty}^{\infty} {\rm d}E \> \bigl(E^2 - (\xi -
  {\textstyle {1 \over 4}})\nabla^2 \bigr) \bigl\{ \Pi_B(E;\vec x\,) -
  \Pi_A(E;\vec x\,) \bigr\} \nonumber \\ 
&+  {1 \over 2}
  \int\limits_{-\infty}^{\infty} {\rm d}t\> \bigl( \lambda {\ddot
  \lambda} - {\dot \lambda}^2 \bigr) \bigl\{ \bra{B}\hat \varphi^2
  (t;\vec x\,)\ket{B} - \bra{A}\hat \varphi^2(t;\vec x\,)\ket{A}
  \bigr\} ,\nonumber\\ 
\end{align}
where $\hat \rho_\xi = - \hat T^t{}_t$ is the energy density operator,
 $\xi$ denotes the coupling to the scalar curvature and the overdot
 represents differentiation with respect to $t$.  It is clear here
 that the relationship between the energy density and the detector
 response depends (not surprisingly) on the rate at which the
 switching is turned on and off.  This statement and that relating to
 $\langle \hat \varphi^2 \rangle$ above are, of course, strongly
 dependent on the particular choice of coupling we have made.  To
 conclude we simply note that in the case of Gaussian switching one
 does obtain a somewhat closer relationship:
\begin{eqnarray}
&& \int\limits_{-\infty}^\infty {\rm d}t \> \lambda^2(t) \bigl\{
  \bra{B}\hat \rho_\xi (\vec x\,)\ket{B} - \bra{A}\hat \rho_\xi (\vec
  x\,)\ket{A} \bigr\} =\nonumber \\ &&{1 \over 2\pi}
  \int\limits_{-\infty}^{\infty} {\rm d}E\> \bigl(E^2 - (\xi -
  {\textstyle {1 \over 4}})\nabla^2 - {\textstyle {1 \over
  2}}T^{-2}\bigr) \times \nonumber
   \\ && \bigl\{ \Pi_B(E;\vec x\,) - \Pi_A(E;\vec x\,)
  \bigr\} .
\end{eqnarray}

\section{PURE VACUUM EFFECTS}
\label{sec:vacuum}
A crucial difference between leaving a detector switched on for all time 
and introducing some form of switching is that switching itself will
induce transitions.  In particular, a switched static detector  moving through a
static space-time in its natural vacuum state will become excited.  This is 
the effect we wish to study in this section.

  In a static space-time we may introduce a complete normalised 
set of mode functions 
of the form ${\rm e}^{-i\Omega t}f_{\vec k}(\vec x\,)$ where 
$\Omega=\Omega(\vec k)$ is positive.  This set may be used to define a natural 
vacuum state $\ket{0}$.
The corresponding vacuum Wightman function at equal spatial points is
\begin{equation}
  \bra{0} \hat\varphi(t,\vec x\,) \hat\varphi(t',\vec x\,) \ket{0}
  = \sum\limits_{\vec k} {\rm e}^{-i\Omega (t-t')}
                  \bigl| f_{\vec k}(\vec x\,)\bigr|^2  . 
\label{eq:3.1}
\end{equation}
Inserting this form into Eq.~(\ref{eq:2.6}), we find that for a static detector at $\vec
x$
\begin{equation}
   \Pi_0(E;\vec x\,) =  \sum\limits_{\vec k} 
   \Bigl|\tilde \lambda\bigl((-g_{tt})^{- {1 \over 2}}\Omega+E\bigr) \Bigr|^2
                 \bigl| f_{\vec k}(\vec x\,)\bigr|^2  . 
\label{eq:3.2}
\end{equation}

\begin{figure}
\includegraphics[width=5.2truecm,angle=270,bbllx=0truecm,bblly=0truecm,bburx=350pt,bbury=600pt]{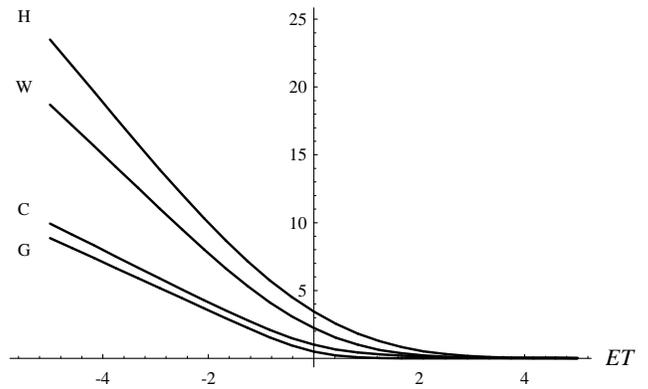}
\caption{\label{fig:vacuum} 
Vacuum response curves for switched detectors.  The letters
$G$, $C$, $W$ and $H$ denote Gaussian switching, Cauchy switching, Welch 
switching and Hanning switching, respectively.}
\end{figure}

   For the Minkowski vacuum Eq.~(\ref{eq:3.2}) takes the form
\begin{eqnarray}
 \Pi_0(E) &=& {1 \over 4 \pi^2} \int\limits_0^\infty k{\rm d}k\> 
     \bigl|\tilde \lambda(k+E) \bigr|^2      \nonumber       \\
 &=& {1 \over 4 \pi^2} \int\limits_0^\infty w{\rm d}w\> 
     \bigl|\tilde \Lambda(w+ET) \bigr|^2        .    
\end{eqnarray}
In the case of sudden switching we have
\begin{equation}
 \Pi_0(E) = {1 \over  \pi^2} \int\limits_0^\infty w{\rm d}w\> 
     {\sin^2 (w+ET) \over (w+ET)^2}  
                                \label{eq:3.5}
\end{equation}
which diverges  (logarithmically) at the upper limit.
For the other switchings introduced in Sec.\ \ref{sec:model} the vacuum responses are finite;
they are  illustrated as functions of $ET$ in Fig.\ \ref{fig:vacuum}.  The region $ET>0$
corresponds to excitation of the detector from its ground state while
the region $ET<0$ corresponds to de-excitation. 
%These graphs are plotted for 
%fixed $T$ and there is no problem as $ET \to 0$ in contrast to the situation 
%when $E$ is fixed and the limit $T \to 0$ is considered.

Having explicitly illustrated the effects of excitation due to switching,
from now on we shall consider the difference between the response in some given 
state $\ket{A}$ containing negative energy density or a negative energy flux
and the vacuum $\ket{0}$:
\begin{equation}
   \Delta \Pi_A(E;\vec x\,) = \Pi_A(E;\vec x\,) - \Pi_0(E;\vec x\,) .  
                       \label{eq:3.6}
\end{equation}
$\Delta\Pi_A$ will be finite even for sudden switching as the high frequency 
divergence is independent of state.

%\begin{center}
\begin{figure}[htb]
\vbox{
\begin{center}
\includegraphics[width=5.2truecm,angle=270,bbllx=0pt,bblly=0pt,bburx=350pt,bbury=600pt]{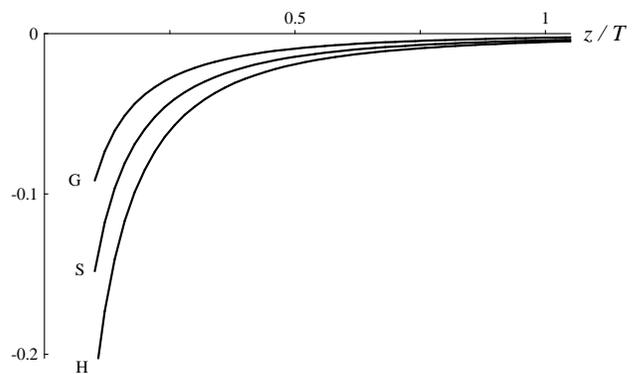}
\end{center}
\caption{\label{fig:casimir} 
Response curves for energy $ET =1$ for a range of switched detectors a
distance $z$ above a single Casimir plate.  The letters $G$, $S$ and
$H$ denote Gaussian switching, Sharp switching and Hanning switching,
respectively.}
}
\end{figure}\noindent
%\end{center} 

    To conclude this section we study a case of a static negative energy
density before turning to negative energy fluxes in the next section.
The simplest configuration to study is the field in its vacuum state
$|{\rm Cas}\rangle$ outside a single Casimir plate $z=0$ on which the
field is taken to vanish.  For this configuration
\begin{equation}
    \langle {\rm Cas}|\hat \varphi^2(z) |{\rm Cas} \rangle    
-   \langle 0 |\hat \varphi^2(z) |0 \rangle  = - {1 \over 16 \pi^2 z^2} ,
                          \label{eq:3.6cas}
\end{equation}
which diverges to $-\infty$ as the plate is approached, and
\begin{equation}
    \langle {\rm Cas}|\hat \rho_\xi(z) |{\rm Cas} \rangle    
-  \langle 0 |\hat \rho_\xi(z) |0 \rangle  = - {1 \over 16 \pi^2 z^4} (1-6\xi) .
                          \label{eq:3.7}
\end{equation}
In these equations $| 0 \rangle$ denotes the standard Minkowski vacuum.
A calculation from Eq.~(\ref{eq:3.2}) gives
\begin{align}
   &\Delta \Pi_{\rm Cas}(E;z) = - {1 \over 4 \pi^2}
       \int\limits_0^\infty k{\rm d}k\> 
     \bigl|\tilde \lambda(k+E) \bigr|^2 \>    
    {\sin (2kz) \over 2kz} \nonumber\\
    &\ =  - {1 \over 4 \pi^2 }
       \int\limits_0^\infty w{\rm d}w\>  
              \bigl|\tilde \Lambda(w+ET) \bigr|^2 \>    
    {\sin \left(2w(z / T)\right)     \over 2w( z / T) } .
\end{align}
The response function $\Delta \Pi_{\rm Cas}(E,\vec x\,)$ is plotted
for our range of switching functions in Fig.~\ref{fig:casimir}.
It is clear from this
that stimulated emission and absorption are reduced by the presence of
the plate. This is a well known and experimentally observed effect.
Note that as $\displaystyle {z / T} \to 0$, $\Delta\Pi_{\rm Cas}(E;z)
\to - \Pi_0(E)$ since in this case the detector cannot become excited
as can also be seen from Eq.~(\ref{eq:3.2}) on noting that in this case
$|f_{\vec k} (z\!=\!0)| =0$.

 As a check on our calculations we may consider the response of an eternal 
detector by taking $\Lambda(\tau) =1$ $\forall \tau$, 
so $\tilde \Lambda(\omega) = 2\pi \delta(\omega)$.  In that case we find
\begin{equation}
   \Delta \Pi_{\rm Cas}(E;z) =   
\begin{cases}
0 & E>0, \\ 
\displaystyle 2 \pi \delta(0) \> 
                {1 \over 4\pi} {\sin (2Ez) \over z} &E<0, \\
\end{cases}
\end{equation}
in agreement with the results of Ref.~\cite{Davies:1989} for the response 
\textsl{per unit time} on identifying $2\pi\delta(0)$ as the total time of the 
measurement.  As expected, in this case energy conservation prohibits 
excitation while de-excitation is affected by the presence of the mirror.

\section{GENERAL STATE WITH ONE MODE EXCITED}
\label{sec:onemode}

   A state of sufficient generality to illustrate the reponse of our switched 
detectors to negative energy fluxes is that of the most general state in which 
just a single mode of momentum $\vec k$ is excited. This may be written as
\begin{equation}
    \ket{\Psi} = \sum\limits_{n=0}^\infty c_n\ket{n} , 
\label{eq:4.1}
\end{equation}
where $\ket{n}$ denotes the $n$th excited state of the particular chosen mode
and $\displaystyle \sum\limits_{n=0}^\infty |c_n|^2 = 1$.
For simplicity we shall use a box normalization with box volume $V$.
Without loss of generality we choose $\vec k = k\hat x$, then
it is straightforward to calculate that
\begin{eqnarray}
    \langle \Psi|\hat \varphi (t, x)\hat \varphi (t',  x') |\Psi \rangle    
-     \langle 0 |\hat \varphi (t,x)\hat \varphi (t',  x')   |0 \rangle
=\nonumber \\
  {1 \over k V }     \biggl\{
          \cos k\bigl((t-t')-(x-x')\bigr) \sum\limits_{n=0}^\infty n|c_n|^2
   + \nonumber \\
    \Re e \Bigl[ {\rm e}^{-ik\left((t+t')-(x+x')\right)} 
     \sum\limits_{n=2}^\infty     \sqrt{n(n-1)}   c_{n-2}^{\>*}c_n  \Bigr]  
               \biggr\} ,
\label{eq:4.2}
\end{eqnarray}             
where $\Re e$ denotes the real part.  It follows that
\begin{align}
  & \Delta \langle \hat \varphi^2 \rangle =
{1 \over k V}      \biggl\{
   \sum\limits_{n=0}^\infty    n|c_n|^2 \nonumber \\
   &\  {} +  \Re e \Bigl[  {\rm e}^{-i2k(t-x)} 
   \sum\limits_{n=2}^\infty    \sqrt{n(n-1)}c_{n-2}^{\>*}c_n \Bigr] \biggr\} ,
                       \label{eq:4.3}             
\end{align}
and
\begin{align}  
   &\Delta \langle \hat \rho_\xi \rangle =
\Delta \langle \hat F_\xi \rangle =
{k \over V }      \biggl\{
              \sum\limits_{n=0}^\infty       n|c_n|^2 \nonumber \\
  &\  {}  + (4\xi -1)  \Re e \Bigl[  {\rm e}^{-i2k(t-x)} 
     \sum\limits_{n=2}^\infty \sqrt{n(n-1)} c_{n-2}^{\>*}c_n \Bigr] \biggr\}.
                       \label{eq:4.4}             
\end{align}
where ${\hat \rho}_\xi = - \hat T^t{}_t$ and $\hat F_\xi = \hat T^x{}_t$,
are the energy density and right-moving energy flux respectively.
All of these expressions correspond to the standard normal-ordered expectation 
values. The cross-terms here enable these quantities (which would
classically be positive definite) to take either sign. 
The  frequency of these cross terms is double that of the 
fundamental mode  highlighting the interference nature of 
negative energy fluxes.

We now turn to our detector response.  A straightforward calculation reveals
\begin{eqnarray}
 &&\Delta\Pi_\Psi (E;x) = \nonumber \\
    && {1 \over 2 kV} 
     \biggl\{     \Bigl( \bigl|\tilde \Lambda(kT+ET) \bigr|^2
+ \bigl|\tilde \Lambda(kT-ET) \bigr|^2 \Bigr) \sum\limits_{n=0}^\infty n|c_n|^2
                \nonumber \\ 
&& {\qquad} + 2 \Re e \Bigl[     \tilde \Lambda(kT+ET) 
         \tilde \Lambda(kT-ET){\rm e}^{i2kx} \nonumber \\ 
      & & {\qquad\qquad} \times
 \sum\limits_{n=2}^\infty \sqrt{n(n-1)} c_{n-2}^{\>*}c_n \Bigr] \biggr\} . 
                       \label{eq:4.5}
\end{eqnarray}
There are a number of observations to make about Eq.~(\ref{eq:4.5}):

(a)~$\Delta \Pi_\Psi(E,\vec x\,)$ is symmetric under $E \to -E$ as follows 
mathematically from the reality of the difference of the two Wightman functions
and physically from  
the relationship between stimulated emission and absorption.
In particular, although Grove's discussion is  expressed purely in terms
of absorption, the approach here is entirely consistent with his results.

(b)~It is easy to check that Eq.~(\ref{eq:2.14}) holds for $\Delta\langle{\hat 
\varphi}^2 \rangle$ of Eq.~(\ref{eq:4.3}) and $\Delta\Pi_\Psi$ of Eq.~(\ref{eq:4.5}) by virtue of 
Parseval's theorem.

(c)~For the special case of an $n$ particle state ($c_n=1$, all others zero)
we have
\begin{equation}
 {\Delta \Pi}_{|n\rangle} (E;x) = 
     {n \over 2 k V} \Bigl\{ \bigl|\tilde \Lambda(ET+kT) \bigr|^2
       + \bigl|\tilde \Lambda(ET-kT) \bigr|^2 \Bigr\} .
\label{eq:npart}
\end{equation}
That the response is proportional to $n$ reassures us that in this simple case 
at least our switched monopole is acting as a particle detector.

(d)~ Using the identity~\cite{Ford:1991}
\begin{eqnarray}
   &&  \left( |\alpha|^2   + |\beta|^2 \right)  \sum\limits_{n=0}^\infty n|c_n|^2
\nonumber \\ &&  +    2 \Re e \left[        \alpha \beta 
 \sum\limits_{n=2}^\infty \sqrt{n(n-1)} c_{n-2}^{\>*}c_n \right] 
         \geq - |\beta|^2 ,       
\label{eq:4.6}
\end{eqnarray}
valid for arbitrary complex numbers $\alpha$, $\beta$ and $c_n$ such that $\displaystyle \sum\limits_{n=0}^\infty |c_n|^2 = 1$ we see that
\begin{equation}
   \Delta\langle{\hat \varphi}^2 \rangle \geq - {1 \over kV}   
\label{eq:4.7}
\end{equation}
and 
\begin{equation}
   \Delta\Pi_\Psi \geq - {1 \over 2kV}   \min 
 \Bigl( \bigl|\tilde \Lambda(kT+ET) \bigr|^2, \bigl|\tilde \Lambda(kT-ET) \bigr|^2 \Bigr).
                                     \label{eq:4.8}
\end{equation}
Thus while either 
$\Delta\langle{\hat \varphi}^2 \rangle$ or  $\Delta\Pi_\Psi$  may be negative
there is a limit as to how negative they can be.
Eq.~(\ref{eq:4.7}) is the direct analogue for $\langle{\hat \varphi}^2
\rangle$ of the results obtained by Ford~\cite{Ford:1991} for
components of $\langle{\hat \rho} \rangle$.  An interesting insight into
Eq~(\ref{eq:4.8}) is given by noting that
\begin{align}
 &{\Delta \Pi}_{|1\rangle} (E;x) = 
     {1 \over 2 k V} \Bigl\{ \bigl|\tilde \Lambda(ET+kT) \bigr|^2
       + \bigl|\tilde \Lambda(ET-kT) \bigr|^2 \Bigr\} \nonumber \\
 &\quad \geq  2 {1 \over 2kV}   \min 
 \Bigl( \bigl|\tilde \Lambda(kT+ET) \bigr|^2, \bigl|\tilde 
               \Lambda(kT-ET) \bigr|^2 \Bigr) .
\end{align}
so that 
\begin{equation}
   \Delta\Pi_\Psi \geq - \frac12 {\Delta \Pi}_{|1\rangle} (E;x) .
\label{eq:onepart}
\end{equation}
This equation admits the following natural semi-classical
interpretation. The detector may
be thought to respond to the zero-point energy we have subtracted 
in forming $\Delta\Pi_\Psi$ exactly as an $n$-particle state with
$n=\frac12$. If we allow for the vacuum fluctuations in this
way the total response will always be positive:
\begin{equation}
   \Delta\Pi_\Psi + {\Delta \Pi}_{|-\frac12 \rangle} (E;x)\geq 0 .
\label{eq:semiclassical}
\end{equation}
where ${\Delta \Pi}_{|-\frac12 \rangle} (E;x)$ is understood to be
formally defined by Eq.~(\ref{eq:npart}) with $n=-\frac12$.

A case of particular interest is that of (single mode) squeezed states
defined for any complex number $\zeta$ by 
\begin{equation}
   \ket{\zeta} = \exp\left[ {\textstyle {1 \over 2}} \zeta^* {\hat a}^2
     - {\textstyle {1 \over 2}} \zeta ({\hat a}^\dagger)^2 \right] 
                                       \ket{0}  . 
\label{eq:4.9}
\end{equation}
This state may be written in the present form with $c_n=0$ for $n$ odd
and
\begin{equation}
    c_{2n} = (\cosh r)^{-1/2} {\bigl[ (2n)!\bigr]^{1/2} \over n!}
 \bigl(- {\textstyle {1 \over 2}} {\rm e}^{i \theta} \tanh r\bigr)^n  ,
                              \label{eq:4.10}
\end{equation}
where $\zeta = r {\rm e}^{i \theta}$.  Correspondingly, we have
\begin{equation}
   \Delta \langle \hat \varphi^2 \rangle =
{1 \over k V}      \biggl\{
 \sinh^2 r -\sinh r \cosh r \cos \bigl[ 2k(x-t)+\theta \bigr] \biggr\} ,
                       \label{eq:4.11}             
\end{equation}
and
\begin{eqnarray}  
  && \Delta \langle \hat \rho_\xi \rangle =
\Delta \langle \hat F_\xi \rangle =
{k \over V }      \biggl\{
  \sinh^2 r - \nonumber \\ && - (4\xi -1) \sinh r \cosh r 
         \cos \bigl[ 2k(x-t)+\theta \bigr] \biggr\} .\qquad
                       \label{eq:4.12}             
\end{eqnarray}
Thus for a fraction $\cos^{-1}(\tanh r)/\pi$ of each cycle, $\Delta
\langle \hat \varphi^2 \rangle$ is negative; this is always less than
half, tending to one half as $r$ tends to infinity.  The average value
of $\Delta \langle \hat \varphi^2 \rangle$ over a cycle is $\sinh^2 r
/(kV)$ which is, of course, positive.  For minimal coupling the energy
density will be negative for an equal time but will be out of phase with
$\Delta \langle \hat \varphi^2 \rangle$.  For other physical choices
of couplings ($0 < \xi \leq 1/6$), the energy density may or may not
be negative depending on the degree of squeezing (magnitude of $r$).
Whenever it is, it will always be out of phase with $\Delta \langle
\hat \varphi^2 \rangle$.  The minimum value of $\Delta \langle \hat
\varphi^2 \rangle$ is 
\begin{equation}
\Delta \langle \hat \varphi^2 \rangle_{\rm min} 
     = - {1 \over kV} \{1 - {\rm e}^{-2r} \} ,   \label{eq:4.13}
\end{equation} 
which is, of course, consistent with the bound (\ref{eq:4.7}).

\begin{figure}[tb]
\vbox{
\includegraphics[width=5.2truecm,angle=270,bbllx=0pt,bblly=0pt,bburx=350pt,bbury=600pt]{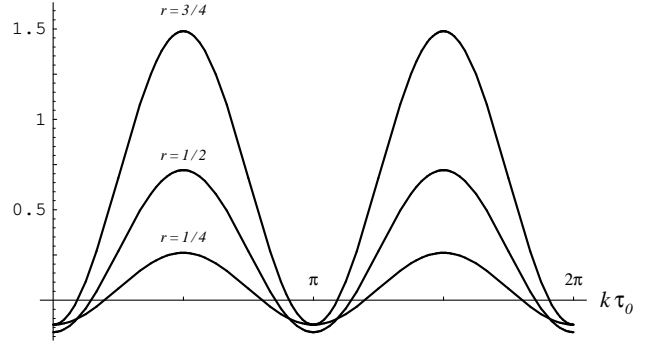}
\caption{
\label{fig:do3} 
Response curves for
squeezed states with squeezing factor  $r={1 \over 4}$, $r={1 \over 2} $ and $r={3 \over 4}$ for a detector with $\chi =1$.}
}
\end{figure}

The detector response to a squeezed state is given by
\begin{align}
 &{\Delta \Pi}_\zeta(E;x) =\nonumber\\ 
    & {1 \over 2 kV}
     \biggl\{     \Bigl( \bigl|\tilde \Lambda(kT+ET) \bigr|^2
   + \bigl|\tilde \Lambda(kT-ET) \bigr|^2 \Bigr) \sinh^2 r \nonumber\\
& - 2 \sinh r \cosh r \ \Re e \Bigl[  
     \tilde \Lambda(kT+ET) \tilde \Lambda(kT-ET){\rm e}^{i(2kx+\theta)} \Bigr] \biggl\},
     \label{eq:4.14}
\end{align}
or, equivalently,             
\begin{align}
&{\Delta \Pi}_\zeta(E;x) =  {1 \over 2 kV}    \Bigl( \bigl|\tilde \Lambda(kT+ET) \bigr|^2
   + \bigl|\tilde \Lambda(kT-ET) \bigr|^2 \Bigr)    \times\nonumber \\
&\ \sinh r \cosh r \bigl\{  \tanh r
 -  \tanh \chi \cos {(2kx+\theta + \phi)}  \bigr\},
\label{eq:4.15}
\end{align}
where 
\begin{equation}
     \tanh\chi(kT,ET) \equiv   
 { 2 \bigl|\tilde \Lambda(kT+ET) \bigr| \bigl| \tilde \Lambda(kT-ET)  \bigr|
    \over   \bigl|\tilde \Lambda(kT+ET) \bigr|^2   + \bigl|\tilde \Lambda(kT-ET) \bigr|^2}
\end{equation}
and $\phi \equiv {\rm Arg} \bigl[\tilde \Lambda(kT+ET) \tilde
\Lambda(kT-ET)\bigr]$.  For the switchings of Sec.\ \ref{sec:model},
which are all symmetric about $\tau=\tau_0$, we have $\phi = -
2k\tau_0$.

It is clear from Eq.(\ref{eq:4.15}) that for fixed $kT$ and $ET$ there
is a critical degree of squeezing, given by $0 < |\zeta| = r < \chi$
required for the squeezed state to produce a suppression of vacuum
excitation.  For fixed $kT$ and $ET$ the minimum value attained by
$\Delta\Pi_\zeta$ occurs for $r = {1 \over 2}\chi$, and in this case,
one finds that for suitably chosen  $x$ (or $\tau_0$) the lower bound (\ref{eq:onepart}) 
is achieved.  Fig.\ \ref{fig:do3} illustrates the response for $\chi=1$ as $k\tau_0$ is 
varied for $r={1 \over 4}$, $r={1 \over 2} $ and $r={3 \over 4}$.

\begin{figure}[tb]
\begin{center}
\vbox{
\includegraphics[width=5.2truecm,angle=270,bbllx=0pt,bblly=0pt,bburx=350pt,bbury=600pt]{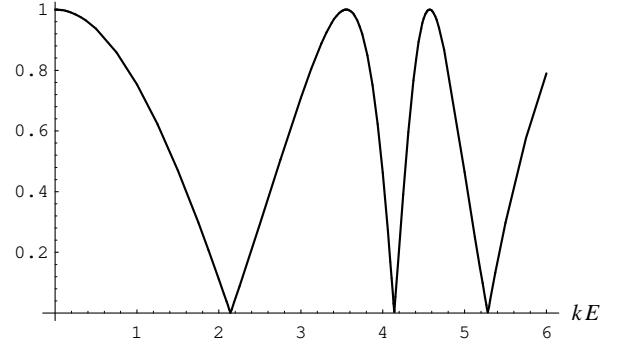}
\caption{
\label{fig:do4} 
$\tanh \chi_S(kT,ET)$ for sharp switching  plotted as
a function of $kE$ for $kT=1$.}
}
\end{center}
\end{figure}

For sharp, Hanning and Welch switching the behaviour of $\tanh \chi(kT,ET)$
as a function of $kT$ and $ET$ is quite complicated; for illustration,
$\tanh \chi_S(1,ET)$ for sharp switching is plotted as a function of
$kE$ in 
Fig.\ \ref{fig:do4}.
For Gaussian and Cauchy switching, $\tanh \chi$ is a monotonic 
decreasing function of $kT$ and $ET$. In the latter cases the explicit 
forms are sufficiently simple to be worth noting, we have
\begin{equation}
    \tanh \chi_G = \sech(2kET^2),
\end{equation}
and 
\begin{equation}
    \tanh\chi_C = 
\begin{cases}
\sech(2ET) & ET < kT,\\ 
\sech(2kT) & ET \geq kT.\\
\end{cases}
\end{equation}

\noindent

Another  simple case in which there is a period of negative energy flux, 
which has been of historical importance, is that of
the vacuum mixed with a two-particle state.  In this case we have
\begin{equation}
     \ket{\Psi} = {1 \over \surd (1+\epsilon^2)} \bigl( \ket{0} + 
          \epsilon \ket{2} \bigr)  ,    \label{eq:4.16)}
\end{equation}
where without loss of generality we have taken $\epsilon$ to be real.  
Corresponding to Eqs.(\ref{eq:4.3}) and (\ref{eq:4.4}) we have
\begin{equation}
\Delta \langle \hat \varphi^2 \rangle = 
          {1 \over kV} {1 \over 1+\epsilon^2} \bigl\{ 2\epsilon^2
           +\sqrt{2}\epsilon \cos2k(t-x) \bigr\} \label{eq:4.17)}
\end{equation}
and
\begin{align}
 \Delta  &\langle \hat\rho_\xi \rangle =  \langle \hat F_\xi  \rangle = \nonumber\\
   &=  {k \over V} {1 \over 1+\epsilon^2} \bigl\{ 2\epsilon^2
           +\sqrt{2}(4\xi-1)\epsilon \cos2k(t-x) \bigr\} .\label{eq:4.18}
\end{align}
Clearly for small $\epsilon$ both $\Delta \langle \hat \varphi^2 \rangle$
and  $\Delta  \langle \hat \rho_\xi \rangle$ can be negative 
for approximately half the time, and, as before,  for physical 
choices of couplings ($0 \leq \xi \leq 1/6$) these times are out of phase.
Corresponding to Eq.(\ref{eq:4.5}) we have
\begin{align}
 &{\Delta \Pi}_\Psi  (E;x) =\nonumber\\ 
     & \quad {1 \over 2\pi k (1 +\epsilon^2)} 
     \biggl\{ \epsilon^2\Bigl(\bigl|\tilde \Lambda(E+k) \bigr|^2
  + \bigl|\tilde \Lambda(E-k) \bigr|^2\Bigr)  \nonumber \\
 &\qquad + \Re e \Bigl[ \sqrt{2} \epsilon
        \tilde \Lambda(kT+ET) \tilde \Lambda(kT-ET) {\rm
     e}^{i2kx}\Bigr] \biggr\} \nonumber \\
   &\quad =  {1 \over 2\pi k (1 +\epsilon^2)} \Bigl(\bigl|\tilde \Lambda(E+k) \bigr|^2
  + \bigl|\tilde \Lambda(E-k) \bigr|^2\Bigr) \times  \nonumber \\
   &\qquad \qquad 
     \Bigl\{ \epsilon^2+ {\epsilon \over \sqrt{2}} \tanh \chi
        \cos(2kx+\phi) \Bigr\} . 
\label{4.19}
\end{align}
It is clear that for $\sqrt{2}|\epsilon| <  \tanh \chi$ we will, as 
before, have  periods when the excitation is less that in the vacuum.

We should add that the similarity of this case to that of squeezed states is 
not accidental: if we work only to order $\epsilon^2$ then the vacuum plus 
two particle state coincides with a squeezed state with $\zeta = - \sqrt{2} 
\epsilon$.  

\section{RINDLER SPACE}
\label{sec:rindler}

\begin{figure}[h]
\begin{center}
\includegraphics[width=5.2truecm]{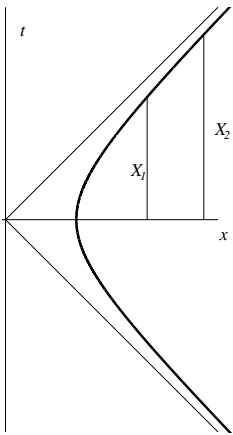}
\caption{
\label{fig:do5} 
The measurement time of an inertial detector in Rindler space $x >
|t|$ is limited by the presence of 
the boundary of Rindler space (mirror) at $x=t$.  Candelas and
Sciama chose to consider the limit $(X,T) \to \infty$ in such a way that 
the final acceleration $A=(X^2 - T^2)^{-1/2}$ remained constant as for
the two trajectories marked $X_1$ and $X_2$ here. }
\end{center}
\end{figure}

  Following Candelas and Sciama~\cite{Candelas:1984} and
 Grove~\cite{Grove:1988}, we will
 now study the  response of an inertial detector moving through the
 Rindler vacuum,  $\ket{R}$, defined in the wedge $x>|t|$ as
 illustrated in Fig.~\ref{fig:do5}.
The Rindler vacuum may be thought of as
the natural vacuum state in the gravitational field of an infinite
flat earth~\cite{Sciama:1981} and is analogous to the Boulware vacuum of
Schwarzschild space-time while the Minkowski vacuum is analogous to
 Hartle-Hawking vacuum. 
This scenario is thus related to the question posed by the title of Candelas and
Sciama's paper~\cite{Candelas:1984}: `Is there a quantum equivalence principle?' in
which the authors addressed teh question of whether a detector falling
 freely in Schwarzschild space-time 
could distinguish if it was moving through the Hartle-Hawking vacuum
or the Boulware vacuum.

An inertial detector moving through the
 Rindler vacuum \textit{must} make a finite time measurement as the detector will
reach the boundary of Rindler space in a finite proper time. 
This boundary plays the role of a mirror in that the field vanishes
 there; indeed the  Rindler vacuum may be realised as the natural
 vacuum above a uniformly accelerating mirror in the limit that the
acceleration tends to infinity.

   Without loss of generality we may take the detector to be at fixed $x$,
$x=X$ say.  Then expressing the Rindler Wightman function in Minkowski 
coordinates we have
\begin{align}
 &\langle R|\hat \varphi (t,X,y,z)\hat \varphi (t',X,y,z) |R \rangle  \nonumber \\    
 & \quad  =  {1 \over 2\pi^2(\eta^2 -{\eta'}^2)} {\displaystyle 
            \ln\left({\eta \over \eta'}\right)   \over \displaystyle 
\left[ \ln^2\left({\eta \over \eta'}\right) -(\tau -\tau')^2 \right]}  \nonumber \\
& \quad  = {1 \over 4\pi^2 (t^2-{t'}^2)} \left[   
   {1 \over \displaystyle  \ln \biggl({X - t' \over X - t}\biggr) }
  + {1   \over \displaystyle \ln \biggl({X + t' \over X +t}\biggr)}
                 \right], 
\label{eq:5.1} 
\end{align}
where, as usual, $t= \eta \sinh \tau$ and $x= \eta \cosh \tau$.
In Eq.~(\ref{eq:5.1}) $t-t'$ is understood to occur in the combination 
$t-t'-i\epsilon$ appropriate to its character as a distribution.

\begin{figure}[h]
\begin{center}
\includegraphics[width=5.2truecm,angle=270,bbllx=0pt,bblly=0pt,bburx=350pt,bbury=600pt]{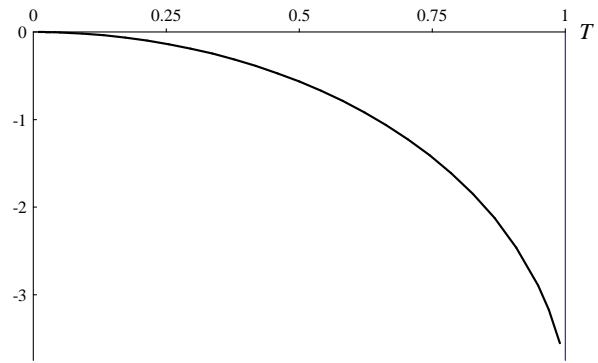}
\caption{
\label{fig:do6} 
Detector response at energy $E=1$ for fixed $X=1$ as $T$ varies.  $T
=1$ corresponds to
reaching the boundary of Rindler space. }
\end{center}
\end{figure}

  For simplicity, we consider a detector switched on 
suddenly  at $t=0$ and  off suddenly at $t= T < X$. 
We have calculated the corresponding response $\Delta\Pi_R$ numerically.
The result taking $X$ fixed and independent of $T$ is plotted in Fig.~\ref{fig:do6}.
That the response for fixed $E$ tends to $- \infty$ as $T \to X$ is to be 
expected on the basis of Eq.~(\ref{eq:2.14}) since  we have 
\begin{equation} 
\Delta \langle \hat \varphi^2 (t)\rangle = 
   - {1 \over 48 \pi^2 \eta^2} = - {1 \over 48 \pi^2 (X^2 - t^2)}    \label{eq:5.2}
\end{equation}
and so 
\begin{equation}
\int\limits_0^T {\rm d}t \> \Delta \langle \hat \varphi^2 (t)\rangle = 
  - {1 \over 96 \pi^2 X} \ln \left({X+T \over X-T} \right) ,     \label{eq:5.3}
\end{equation}
which diverges logarithmically to $-\infty$ as $T \to X$.

\begin{figure}[tb]
\begin{center}
\includegraphics[width=5.2truecm,angle=270,bbllx=0pt,bblly=0pt,bburx=350pt,bbury=600pt]{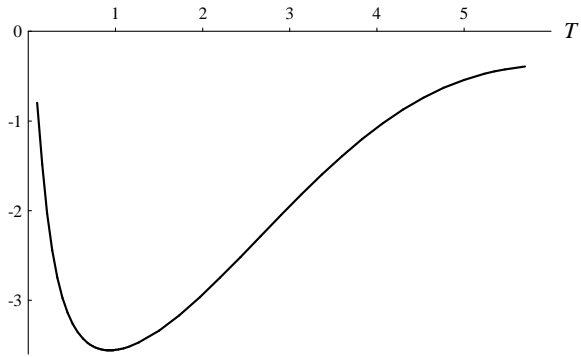}
\caption{
\label{fig:do7}
 Detector response 
at energy $E=1$  as $T$\ varies with $X$ varying so that the final 
acceleration $A=(X^2 - T^2)^{-1/2} $ is held fixed at $7{\cdot}09$
(so $X = 1$ when $T=0{\cdot}99$).}
\end{center}
\end{figure}

Rather than consider the limit illustrated in Fig.~\ref{fig:do6}, Candelas and
Sciama chose to consider the limit $(X,T) \to \infty$ in such a way that 
the final acceleration $A=(X^2 - T^2)^{-1/2}$ remained constant.
The numerically calculated detector response,  $\Delta\Pi_R$,
corresponding to this configuration is plotted in Fig.~\ref{fig:do7}.
The fact that this response tends to zero as  $T \to \infty$ is the 
essence of the result obtained by Candelas and Sciama.  

Both Figs.~\ref{fig:do6} and ~\ref{fig:do7} bear out the conclusion of Grove that as a detector 
approaches the mirror the reduction in vacuum fluctuations near the mirror 
lead to a sharp reduction in the level of excitation of the detector.  This
interesting effect is lost in the limit taken by Candelas and Sciama.

In fact, as  Candelas and Sciama did not subtract the infinite vacuum
excitation introduced by their   switching they were forced to 
consider the time derivative $\displaystyle {{\rm d} \over {\rm d}T} \Pi_R(E)$.
This provides a notion of the difference in response between one ensemble of 
detectors switched on at time 0 and off at time $T$ and a different
ensemble switched on at time 0 and off at time $T+{\rm d}T$.
Grove considered $\displaystyle {{\rm d} \over {\rm d}T} \bigl(\Pi_R(E) -
\Pi_0(E) \bigr)$ but incorrectly asserted that 
$\displaystyle {{\rm d} \over {\rm d}T} \Pi_0(E) = 0$ whereas in reality 
\begin{equation}
{{\rm d} \over {\rm d}T} \Pi_0(E) = {1 \over 2\pi^2} \left[
{\cos ET \over T} + E\,{\rm si}(ET) \right],                  \label{eq:5.4}
\end{equation}
where ${\rm si}(x)$ is the sine integral defined by~\cite{Gradsteyn:1980}
\[
    {\rm si}(x)  \equiv \int\limits_\infty^x {\rm d}t \>{\sin t \over t} .
\]
Eq.~(\ref{eq:5.4}) may be derived either from differentiating
Eq.~(\ref{eq:3.5}) or directly by deforming the contour 
of integration to that used by Candelas and Sciama.  Note that 
\begin{equation}
{{\rm d} \over {\rm d}T} \Pi_0(E) \to  - {E \over 2\pi} \theta(-E)
  \qquad {\rm as} \quad T \to \infty
\end{equation}
as required. Grove's oversight does not in any case effect the analysis 
of  the divergence 
 in  $\displaystyle {{\rm d} \over {\rm d}T} \Pi_R(E)$ as $T \to X$, 
since although $\displaystyle {{\rm d} \over {\rm d}T} \Pi_0(E)$ is  
non-vanishing it is  manifestly regular in this limit.  We shall work with
$\displaystyle {{\rm d} \over {\rm d}T} \bigl(\Pi_R(E) -\Pi_0(E) \bigr) =
{{\rm d} \over {\rm d}T} \Delta\Pi_R(E) $
as this is more natural within our formalism.

Taking account of the foregoing comments, Candelas and Sciama prove that
\begin{align}
{{\rm d} \over {\rm d}T} \Delta\Pi_R(E) \sim 
   -  & {1 \over 8 \pi^2} {\ln\bigl|2 A^2 T/ E \bigr| \over T}
   \nonumber \\
 &\qquad \text{as}  \quad \bigl|A^2 T/E\bigr| \to \infty.                  \label{eq:5.5}
\end{align}

From Eq.~(\ref{eq:2.14}) we immediately obtain
\begin{equation}
     \bra{R} : \hat \varphi^2(T) : \ket{R} =
  {1 \over 2 \pi} \int\limits_{-\infty}^\infty {\rm d}E \>
              {{\rm d} \over {\rm d}T} \Delta\Pi_R(E).    \label{eq:5.6}
\end{equation}
We may follow Grove and use the asymptotic expression (\ref{eq:5.5}) of
$\displaystyle {{\rm d} \over {\rm d}T} \Delta\Pi_R(E)$ for 
$\bigl|E/(2A^2T)\bigr| \leq {\cal O}(1)$ and approximate it as zero for
$\bigl|E/(2A^2T)\bigr| > {\cal O}(1)$.
Then 
\begin{align}
   {1 \over 2 \pi} &\int\limits_{-\infty}^\infty {\rm d}E \>
              {{\rm d} \over {\rm d}T} \Delta\Pi_R(E) \nonumber \\
         &\sim - {1 \over 16\pi^3 T}
        \int\limits_{-{\cal O}\bigl( 2A^2T \bigr)}
                 ^{{\cal O}\bigl( 2A^2T \bigr)}
               {\rm d}E \>   \ln \bigl|E/(2A^2T)|\nonumber \\
         &= - {A^2 \over 4\pi^3 } \int\limits_0^{{\cal O}( 1)}
               {\rm d}x \>   \ln x = - {A^2 \over 4\pi^3 }  {\cal O}(1).
\label{eq:5.7}
\end{align}
Given the crudeness of the calculation the agreement with the exact result
\begin{equation}
   \bra{R} : \hat \varphi^2(T)  : \ket{R} \sim - {A^2 \over 48 \pi^2}
\end{equation}    
is remarkable.  
%Note that we have started here from an exact equation 
%(Eq.(5.6)) while Grove uses a four dimensional generalisation of his two 
%dimensional result which he makes no attempt to justify and which is almost 
%certainly not exact while possessing some element of truth. 

\section{CONCLUSION}
\label{sec:conc}

With our particular choice of linear coupling we have seen the very close link 
between detector response and reduced vacuum noise.   The absence of vacuum 
fluctuations leads to a reduction in the level of excitations of a switched 
detector over that which would have occurred in the vacuum as a result of the 
switching. We may translate this into thermodynamic terms.   We consider a 
hot ensemble of two level atoms which is initially at inverse temperature 
$\beta$ and is then allowed to interact for a finite time with a state 
$\ket{\Psi}$. The ensemble will, of course, cool (lose entropy) if it
is placed just in the vacuum so we consider
the change in entropy relative to the change in the vacuum which is given by
\begin{equation}
   \Delta S = { \beta E \over 1 + {\rm e}^{-\beta E}}
\bigl[  \Delta\Pi_\Psi(E) - {\rm e}^{-\beta E} \Delta\Pi_\Psi(-E) 
\bigr] .
                                                           \label{eq:6.1}
 \end{equation}
Considering for simplicity states for which 
$\Delta\Pi_\Psi(E) = \Delta\Pi_\Psi(-E)$ (for example, squeezed states) we have
\begin{equation}
\Delta S =  \beta E \,{1 - {\rm e}^{-\beta E}\over 1 + {\rm e}^{-\beta E}}\,
                                       \Delta\Pi_\Psi(E)    .     \label{eq:6.2}
\end{equation}
Here the prefactor is manifestly positive so the ensemble which interacted 
with that state $\ket{\Psi}$ will have cooled more than an identical ensemble 
in the vacuum if and only if $\Delta\Pi_\Psi(E) < 0$.

The foregoing results serve to clarify the response of matter to
pulses of negative energy flux of limited duration.  They are broadly 
in accordance with one's intuition that negative energy should have
the effect of enhancing de-excitation, i.e. to induce `cooling'.
However, our results are necessarily somewhat model dependent and for
our standard monopole model we find that there is not always a simple 
relationship between the strength of the negative energy flux and the
behaviour of matter.

Considerable interest attaches to the thermodynamics of negative
energy.  If a sustained negative energy flux could be directed at a
hot body (or a black hole) in such a way as to reduce its temperature,
 hence entropy, by a macroscopic amount there would appear to be a
clear violation of the second law of thermodynamics.  There is a
considerable literature on this topic already.  The results of this
paper are a first step to investigating the thermodynamics of negative
energy.  However, the `cooling' effects we have discussed cannot be
immediately used to draw thermodynamic conclusions, because they have
been restricted to first order in perturbation theory and, as shown by 
Grove~\cite{Grove:1986}, a proper investigation of the thermodynamic implications
necessitates a calculation to second order in perturbation theory.
(At first order alone, it is not possible to determine whether the 
de-excitation effects are merely due to the (small) violation of
energy conservation expected in any process in which a general quantum
state collapses to an energy eigenstate, or whether they pressage a
systematic reduction in the energy of the matter which would have
serious thermodynamic implications.) We shall report on this further
investigation in a separate paper.

\end{document}